\documentclass[
superscriptaddress,
 amsmath,amssymb,
 aps,
pra,
twocolumn,
nofootinbib,
]{revtex4-2}

\usepackage{xcolor}

\usepackage{graphicx}
\usepackage{dcolumn}
\usepackage{bm}
\usepackage{hyperref}

\usepackage{amsmath}
\usepackage{amssymb}
\usepackage{stackrel}
\usepackage[cm]{fullpage}
\usepackage{amsthm}
\usepackage{mathtools}
\usepackage{enumitem}
\usepackage{bbold}
\usepackage[normalem]{ulem}
\usepackage{hyperref}
\usepackage{cleveref}
\usepackage{appendix}
\usepackage{comment}
\usepackage{subcaption}
\usepackage{centernot}
\usepackage{braket}
\usepackage{apptools}
\AtAppendix{\counterwithin{Proposition}{section}}
\usepackage{graphicx}
\usepackage{multirow}
\usepackage{array}
\usepackage{xcolor}
\usepackage{ragged2e}

\theoremstyle{definition}

\usepackage{graphicx}

\newcommand{\blk}{\color{black}}

\begin{document}

%
%
	
\title{Experimental Test of the Principle of Tomographic Locality}
\author{Tristan S. Lismer}
\email{tlismer@uwaterloo.ca}
\affiliation{Institute for Quantum Computing and Department of Physics \& Astronomy,
University of Waterloo, Waterloo, Ontario N2L 3G1, Canada}
\author{Kaleb B. Felefele}
\affiliation{Institute for Quantum Computing and Department of Physics \& Astronomy,
University of Waterloo, Waterloo, Ontario N2L 3G1, Canada}
\author{Robert W. Spekkens}
\email{rspekkens@perimeterinstitute.ca}
\affiliation{Perimeter Institute for Theoretical Physics, 31 Caroline Street North, Waterloo, Ontario Canada N2L, 2Y5}
\author{Kevin J. Resch}
\affiliation{Institute for Quantum Computing and Department of Physics \& Astronomy,
University of Waterloo, Waterloo, Ontario N2L 3G1, Canada}

\begin{abstract}
The principle of tomographic locality states that the operational state of a multipartite system can be fully characterized by the statistics obtained from measurements that are local to 
 the individual subsystems.  This property holds in 
 quantum theory and features prominently in axiomatic reconstructions of the theory, where it serves to rule out a wide class of alternatives. For instance, quantum theory with Hilbert spaces defined over the real field (rather than the complex field) is an example of a theory that is ruled out in this fashion. 
 Given its foundational importance, it is worthwhile to subject this principle to a direct experimental test. Specifically, we consider an experiment on the polarization degrees of freedom of a pair of photonic modes in a prepare-and-measure scenario and analyze the resulting data within the framework of generalized probabilistic theories. The signature of a failure of tomographic locality is that there are pairs of states on the bipartite system that can only be distinguished by the statistics they yield for non-separable measurements. In the full quantum setting, we find no evidence of a violation of tomographic locality. As a test of our analysis method, we also verify that if we restrict attention to those states and measurements that lie within the fragment described by quantum theory over the real field, then a clear signature of  the failure of tomographic locality 
 is observed.
\end{abstract}

\maketitle

\section{Introduction}

In the formalism of generalized probabilistic theories (GPTs)~\cite{hardy2001quantum,barrett2007information,Chiribella2010purification,muller2021probabilistic},
 every preparation of a system is associated with an {\em operational state}, a vector in a real vector space that specifies the statistics for all possible measurements on that preparation procedure. If, in a given theory, the operational state of a composite system can be fully described by the statistics it predicts for measurements that are local to the subsystems, then the theory is said to obey the principle of \textit{tomographic locality}.  If one conceptualizes quantum theory as a GPT, the operational state is simply the density operator and the principle is found to hold.  For instance, a two-qubit quantum state can be written as a sum of products of Pauli operators (where identity is included in the set), so that learning the expectation values of all 16 such products via local measurements allows one to learn the state. 
Tomographic locality is exploited routinely in practical schemes for implementing state tomography in quantum experiments. 

In the field of quantum foundations, there has been a great deal of interest in identifying clear sets of axioms from which the formalism of quantum theory can be derived,
and tomographic locality serves as a central axiom in the most prominent attempts to do so~\cite{hardy2013reconstructing,barrett2007information,Chiribella2010purification,muller2021probabilistic}. 
  The axiomatic program has also promoted the {\em methodology of foil theories}, wherein 
  alternatives to quantum theory are studied not as empirical competitors but for the the insight they yield into what is distinctive about quantum theory.  For example, we can better understand the principle of tomographic locality by studying theories that violate it~\cite{centeno2024twirled}.  The most prominent example of such a theory is real-amplitude quantum theory (wherein the Hilbert space vectors are restricted to those having real components in some basis)~\cite{stueckelberg1960quantum}, as noted in Refs.~\cite{araki1980characterization,wootters1990local,hardy2001quantum}.   Other examples of such theories include quaternionic quantum theory~\cite{hardy2001quantum}, fermionic quantum theory~\cite{darianoFermionic2014,Dariano2014feynmanproblem}, exotic variants of classical theories~\cite{d2020classicality,scandolo2019information,Chiribella2024}, and a post-quantum theory constructed from Euclidean Jordan algebras~\cite{barnum2020composites}.

   Despite its foundational significance, 
   there has not previously been a direct experimental test of the principle of tomographic locality.
   Such a test demands a means of analyzing the experimental data that does not presume the correctness of operational quantum theory, since the assumption of tomographic locality is built into the structure of quantum mechanics itself.  
The framework of GPTs is such a tool.  It allows one to fit experimental data in a theory-agnostic way and without a prior characterization of any of the experimental procedures, as demonstrated in Refs.~\cite{mazurek2021experimentally,grabowecky2022experimentally}.  The scheme described in these works, termed {\em GPT tomography}, makes minimal assumptions about the experiment. 
First, it must be possible to treat the different components of a circuit as independent of the others, in the sense that each can be varied while the others remain fixed. For instance, it should be possible to vary the preparation procedure  independently of the measurement procedure. 
Second, it must be possible to treat the runs of the experiment as 
independent and identically distributed (i.i.d.).
We make both of these assumptions in our experiment.  (Although considering the possibility of these assumptions failing is a worthwhile endeavour,  it is not our concern in this work.)
The technique of GPT tomography makes no assumption about the dimensionality of the GPT vector space describing the states and effects of a system; an estimate of the dimensionality instead emerges from the analysis.


Here we show how GPT tomography provides a means of testing the principle of tomographic locality through a quantum optics experiment on the polarization degrees of freedom of a pair of photonic modes.


We also benchmark our data analysis technique by applying it to an experiment wherein we artificially restrict the preparations and measurements in such a way that we expect to see the signature of a failure of tomographic locality. Specifically, we will not only do the analysis on the full data set but also on the subset corresponding to those preparation and measurement procedures where we are targeting states and effects that are
confined to the real-amplitude sector of quantum theory.  This analysis is meant to simulate what would occur in our experiment if nature were governed by real-amplitude quantum theory rather than complex-amplitude quantum theory.

 

We begin by describing a scheme for testing tomographic locality in a generic experiment.  Further on, we describe the particular experiment we conducted, the details of the data analysis technique, what one expects to see if quantum theory is correct, and the results of the data analysis.

\section{GPT tomography}

The GPT framework represents states on a system $A$ by vectors embedded in a real vector space $V_A$, and effects  by vectors in the dual vector space $V_A^*$ (an effect is a specification of a measurement and an outcome).  We will here assume $V_A$ to be a space of column vectors and $V_A^*$ to be a space of row vectors, with the inner product given by matrix multiplication.  States and effects of a composite system are represented by vectors in the {\em tensor product} of the vector spaces associated to the subsystems. Thus, for a system composed of two parts, $A$ and $B$, 
  $V= V_A \otimes V_B$.  Each preparation procedure on the composite system $AB$ is associated to a vector $\bm{s} \in V_A \otimes V_B$, termed a {\em GPT state vector}, and each effect on $AB$ is associated to a vector $\bm{e} \in V^*_A \otimes V^*_B$, termed a {\em GPT effect vector}.  A transformation on $AB$ is associated to a linear operator $T: V_A \otimes V_B \rightarrow V_A \otimes V_B$, which we shall represent by a matrix.  The probability of registering the outcome associated to the GPT effect vector $\bm{e}$ in an experiment with a preparation associated to the GPT state vector $\bm{s}$ and transformation matrix $T$ is $\bm{e} \cdot T \cdot \bm{s}$, where $\cdot$ represents matrix multiplication. The demand that this expression define a genuine probability is what constrains the choices of the state space, effect space, and space of transformations. For instance, a valid GPT effect vector must satisfy $\mathbb{0} \le \bm{e}
  \le 
 \bm{u}
  $ where the vector $\bm{u}$,
  termed the {\em unit effect}, assigns probability 1 to all states and where the vector $\textbb{0}$,
  termed the {\em zero effect}, assigns probability 0 to all states.



It is clear that if a set of GPT states can be realized, then any convex mixture of these is also realizable.  Thus, the space of GPT states that are realizable in a given experiment is necessarily a convex set. A similar condition holds for GPT effects.


Let $i \in \{1, \dots, n\}$ be an index that ranges over the preparation procedures in an experiment, and $j\in \{1, \dots, m\}$ an index that ranges over the binary-outcome measurements. Let $D$ denote the matrix for which the component $D_{ij}$ is the relative frequency with which one obtained the positive outcome in the $j$th measurement when acting on the $i$th preparation. $D$ constitutes the raw data that results from the experiment.

The first step in the analysis is to find a GPT model $M$ of the experiment such that the matrix $D^{M}$ of probabilities predicted by this model provides a good fit to the matrix $D$  of experimental relative frequencies.

We recall the scheme of GPT tomography for a prepare-measure experiment~\cite{mazurek2021experimentally}.  For each possible choice of dimension $k$ of the GPT vector space $V$, one defines a model $M$ that consists of a set of GPT state vectors in $V = \mathbb{R}^k$, denoted $\mathcal{S} = \{\bm{s}_i\}_{i=1}^{n}$, and a set of GPT effects vectors in the dual space $V^*$, denoted  $\mathcal{E} =  \{\bm{e}_j\}_{j=1}^{m}$. 
 Each such model specifies the probability that the state $\bm{s}_i$ passes the test associated to the effect $\bm{e}_j$ as the inner product $\bm{s}_i\cdot \bm{e}_{j}$.  These probabilities can be organized into a matrix $D^{M}$ with components $(D^{M})_{ij}= \bm{s}_i\cdot \bm{e}_{j}$. 
Among such models, one seeks to find the one that  provides the best-fit to  the experimental data $D$ in the sense that $D^{M}$
  minimizes the $\chi^2$ distance to $D$,
  i.e., minimizes
$\chi^2 = \sum_{ij} \left( \frac{D_{ij} - (D^M)_{ij}}{\Delta D_{ij}} \right)^2$.
\footnote{It is worth noting that there is a freedom in the solution to this optimization problem.  By decomposing $D^{M}$ as $D^{M} = SE$ where $S$ is a $n \times k$ matrix and $E$ is a $k\times m$ matrix, the rows of $S$ are the vectors $\{\bm{s}_i\}_{i=1}^{n}$, and the columns of $E$ are the vectors $\{\bm{e}_j\}_{j=1}^{m}$.  
But then it is clear that such a 
decomposition of $D^{M}$ 
 is not unique: for any invertible $k\times k$ matrix $\Lambda$, one can write 
 $D^M= \blk
 S\Lambda \Lambda^{-1}E$, and consequently there is a linear freedom in the GPT state and effect vectors that yield the predictions in $D^{M}$ and it is purely conventional which choice is made.}


The optimal choice of dimension $k$ of the GPT vector space of the model  is not something that is assumed a priori in the analysis but is rather estimated from the experimental data based on predictive power.  Ruling out dimensions that are too small is straightforward because these lead to models that underfit the data, something that can be identified from the $\chi^2$.  Above a certain dimension of model, however, one expects to always find 
a reasonable $\chi^2$ for the best-fit model.  Nonetheless, it is possible to adjudicate between these dimensions, because models of higher dimension can have less predictive power than models of smaller dimension by virtue of the former {\em overfitting} the data relative to the latter. We estimate the predictive power of different dimensions using a train-and-test methodology, adapting the approach described in Ref.~\cite{daley2022experimentally}



Specifically, we separate the experimental data for each configuration of the experiment into two sets, termed the {\em training set } and the {\em test set} respectively. 
We opted for the training set to consist of 
a random  sample of 90\% of the runs in a given configuration, with the test set consisting of the 
 other 10\% of the runs.
 The matrix of relative frequencies one obtains for each case are denoted $D^\text{train}$ and $D^\text{test}$ respectively.

We define the training error achieved by a model $M$ to be 
$(\chi^{2})^\text{train}_M = \sum_{ij} \left(\frac{D_{ij}^\text{train} - D_{ij}^M 
}{\Delta D_{ij}^\text{train}}\right)^2$, and the test error to be
 $(\chi^{2})^\text{test}_M = \sum_{ij} \left(\frac{D_{ij}^\text{test} - D_{ij}^M 
}{\Delta D_{ij}^\text{test}}\right)^2$. 
The signature of a model $M'$ {\em overfitting} the data relative to a model $M$ is if $M'$ has a smaller training error  than $M$ but a larger test error. 
The most predictive model is the one that minimizes the test error.  In our data analysis, we seek to identify the dimension of the model that is most predictive.




We seeded all of the real parameters with arbitrary initial values and performed the optimization using the SLSQP algorithm implemented through the Python library SciPy~\cite{virtanen2020scipy}.


 \blk

Tomographic locality concerns the tomographic power of states that are of product form, i.e., those that we model theoretically by GPT vectors that factorize across the tensor product structure defining the bipartition of the composite system.  
Put another way, tomographic locality is only meaningful relative to a notion of locality, which is captured in a GPT by a tensor product structure. In our implementation, we target product states and target measurements, but we do not assume that these are achieved perfectly. However, because the transformations we implement 
on each subsystem are physically separated, it is reasonable to model them as independent.



Our answer to the question of how to define the tensor product structure 
will be presented in Sec.~\ref{sec:TPS} and is one of several ways in which the current work advances the framework of GPT tomography introduced in Refs.~\cite{mazurek2021experimentally} and \cite{grabowecky2022experimentally}.






\section{Experiment}

\subsection{Theoretical considerations for the experimental design}\label{3a}




In GPT tomography~\cite{mazurek2021experimentally}, the dimension of the GPT vector space is not assumed a priori, but is an output of the data analysis~\footnote{A difference is that in Ref.~\cite{mazurek2021experimentally}, overfitting was diagnosed using Akaike information criterion rather than via a train-and-test methodology as we do here.}. Any conclusion about dimension, however, rests on the assumption that the realized set of preparation and measurement procedures for the system under investigation spans 
 the full GPT state and effect space.
In other words, one can be confident about the conclusions regarding GPT vector space dimension only to the extent that one is confident that the experiment has probed a set of preparations and measurements that are {\em tomographically complete} for the system under investigation. 
If the quantum prediction about the GPT dimension for the polarizations of a pair of photonic modes 
is incorrect and the true dimension is larger than the quantum prediction, then this failure of quantum theory might only be pronounced in exotic experimental set-ups, where these predictions have not previously been tested. But the failure can also manifest itself in conventional experimental set-ups that are probed with a higher precision than previously achieved.  This is because if the states probed  
{\em do} have a small nonzero component in the post-quantum directions of the GPT vector space, then this component will be \blk 
detectable at sufficiently high precision.   This approach---of looking for deviations from quantum theory {\em at the precision frontier}---is the one taken in Refs.~\cite{mazurek2021experimentally} and \cite{grabowecky2022experimentally} and also in this work.



 

The present article represents the first application of the technique of GPT tomography to a system that is described in quantum theory as a {\em pair} of qubits, 
 which is the simplest case in which one can probe the tensor product structure.  
Earlier works  considered a system that is described in quantum theory as a single qubit~\cite{mazurek2021experimentally} or as a single qutrit~\cite{grabowecky2022experimentally}.

These earlier works did not only look for {\em dimensional} deviations from the predictions of quantum theory.  They also aimed to reconstruct the {\em shapes} of the state and effect spaces 
and thereby to put bounds on the extent of possible deviations from the predictions of quantum theory in these shapes.

The latter goal motivated the particular choice of the set of preparations and measurements that were implemented in those works.
Specifically, there was a focus on ensuring that these sets gave a good discrete approximation to the true shape of the full state and effect spaces. Because these shapes are smooth in the quantum case, there was an emphasis on having a dense set of states and effects.  For instance, Ref.~\cite{mazurek2021experimentally} sought to identify polytopes that provided good inner and outer approximations to the Bloch sphere.  


Insofar as the present work has the goal of testing tomographic locality, it is concerned only with the possibility of a dimensional deviation from the predictions of quantum theory, and       not with deviations of the {\em shapes} of the state and effect spaces.    
Given this shift in emphasis, we are not concerned with obtaining a discrete approximation of the full set of states and effects.

It is therefore sufficient to target a set of quantum states and quantum effects that span the operator space and that are not easily mistaken for spanning  some subspace thereof. It is also optimal if this set includes elements which are highly entangled, since these are the states for which 
 there is the greatest potential to see a failure of tomographic locality. 
  The set must also  include products states wherein the local states on each subsystem span the GPT vector space of that subsystem, and similarly for product effects. 

Finally, recalling that we also aim to test our data analysis technique on a subset of preparations and measurements which in their quantum description are confined to the real-amplitude sector, it is useful to choose a set of quantum states and effects which include real-amplitude subsets with the sorts of properties just mentioned.

The states and effects defined by the {\em stabilizer formalism} for quantum computation~\cite{gottesman1998heisenberg}, which are often used in quantum tomography schemes~\cite{mazurek2021experimentally} fit these desiderata nicely and thus are the ones we target in our experiment. 

Note that there are 60 two-qubit stabilizer states in total, of which 36 are product states (formed from the 6 single-qubit stabilizer states per qubit) and 24 are maximally entangled. Of the 60 two-qubit stabilizer states, 24 lie in the real-amplitude sector, with 16 of them being of product form and 8 being maximally entangled. We refer the reader to the Supplemental Material for the definitions of these states.

\blk

 \subsection{Experimental set-up}

\begin{figure*}[ht]
    \centering
	\includegraphics[width=.7\linewidth]{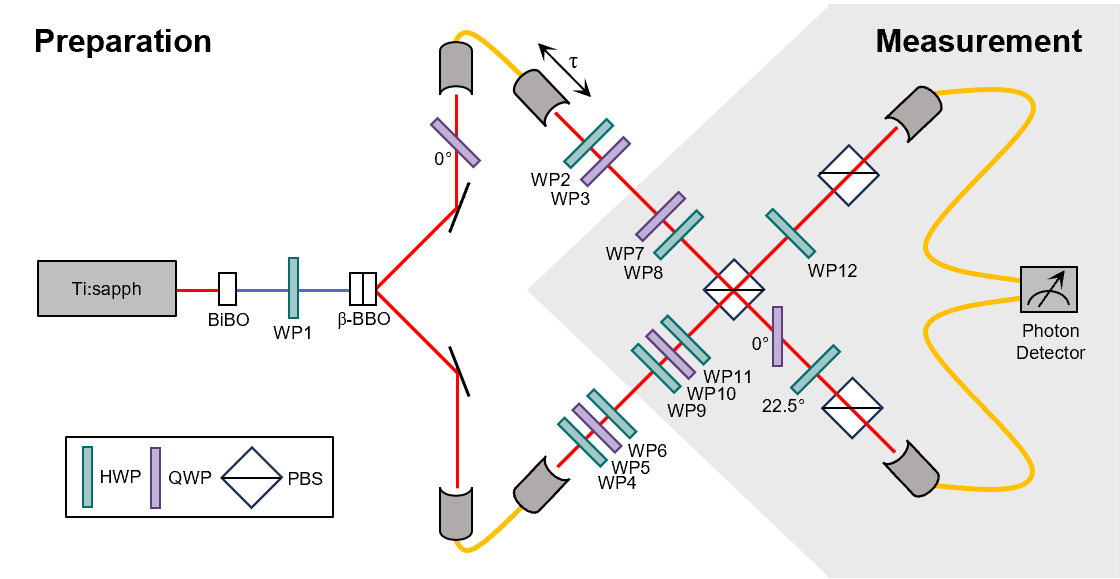}
	
	\caption{\justifying 
    \label{fig:setup}
	  The experimental configuration. A pulsed Ti:sapphire laser (80~MHz repetition rate, 2.4~W average power) centered at 790~nm produces a beam with an average power of 0.7~W at 395~nm through second-harmonic generation via a BiBO (Bismuth borate) crystal. This pump beam generates photon pairs in two orthogonally oriented $\beta$-BBO (barium borate) crystals via type-I down-conversion. The pump polarization is adjusted using the waveplate ${\rm WP}_1$. By setting ${\rm WP}_1$ to $0^\circ$, the $ |H\rangle|H\rangle$ state is prepared, while setting it to $22.5^\circ$ produces the $\frac{1}{\sqrt2}(|H\rangle|H\rangle+|V\rangle|V\rangle)$ state. From these initial states, arbitrary two-photon polarization states can be achieved by controlling the angles of the waveplates ${\rm WP}_2, \dots, {\rm WP}_6$. The measurement section of the setup consists of passing the photons through ${\rm WP}_8, \dots, {\rm WP}_{11}$, before recombining them at a polarizing beam splitter (PBS). An adjustable delay ($\tau$) is introduced in one arm in order to ensure the temporal overlap of the photons. The outputs of the PBS are further processed by a HWP and a PBS in each arm. This process enables the projection onto either the product basis (with ${\rm WP}_{12}$ to $0^\circ$) or the entangled basis (with ${\rm WP}_{12}$ to $22.5^\circ$). To alter the local phases in the preparation and measurement states, a tilted QWP is placed in each section of the setup.}
\end{figure*}

The experimental setup is presented in Fig.~\ref{fig:setup}.  This configuration is designed to enable high-contrast measurements of the targeted states by exploiting the two-photon interference at the PBS in a manner similar to the Hong-Ou-Mandel (HOM) effect. 

To obtain HOM-like interference, we place one of the fiber couplers on a translation stage to vary the temporal delay between the photons and ensure their overlap. Photons are counted using single-photon detectors (Perkin-Elmer SPCM-AQ4C), with coincidences being registered using a (UQD) logic unit with a $3\ \text{ns}$ window. The HOM-like interference is observed as a dip in the coincidence rate as a function of the relative temporal delay. With the input polarization state $\frac{1}{\sqrt2}(|H\rangle|H\rangle+|V\rangle|V\rangle)$ and measurement effect
$\frac{1}{\sqrt2}(|H\rangle|H\rangle-|V\rangle|V\rangle)$ optimized for indistinguishability at the PBS, we observe a visibility of 0.94.

Once this dip in coincidences is as small as possible, we proceed to the data collection. From a list of 60 preparation waveplate settings and 60 measurement waveplate settings, we cycle through all 3600 combinations. For each of these, we collect up to 10000 coincidence counts per 3 seconds. These counts are converted into relative frequencies, forming the $60 \times 60$ matrix, denoted by $D$, which is the raw dataset used in the ensuing analysis of
the validity of tomographic locality in quantum theory.

As noted above, of the 60 stabilizer states, 24 lie entirely within the real-amplitude sector. We can therefore extract from $D$ a $24 \times 24$ submatrix of relative frequencies corresponding to the cases where both the state and the effect being targeted \blk are in this real sector. We denote this submatrix by $D^{\rm real}$. This submatrix serves as the starting point for benchmarking our data analysis technique by looking for signatures of the failure of tomographic locality under a restriction to  
the real-amplitude sector of quantum theory.

\begin{figure}[h]
    \centering
	\includegraphics[width=.9\columnwidth]{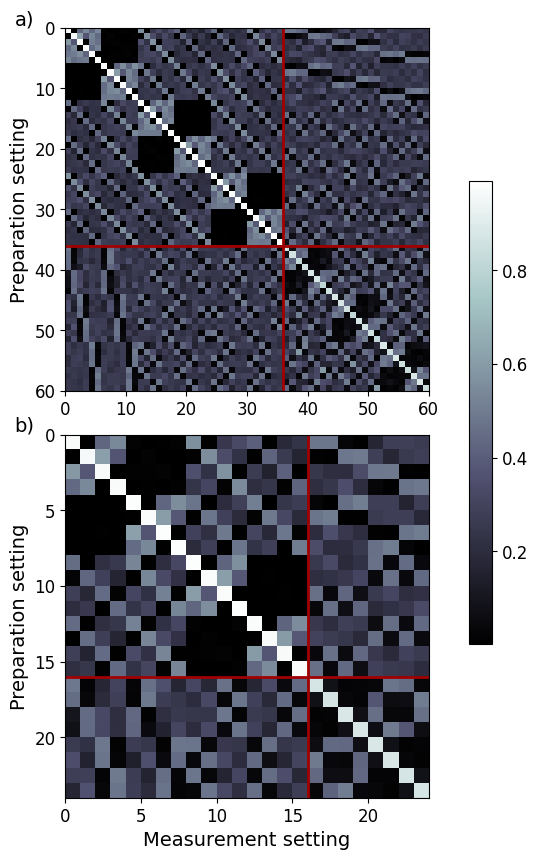}
	
	\caption{\justifying 
    \label{fig:freqMatrix}
 	Experimental data. Measured probabilities for each preparation and measurement setting for matrices a) $D$, and b) $D^{\text{real}}$. The upper-left quadrants correspond to measurements performed with product states and product measurements, while the lower-right quadrants correspond to measurements performed with maximally entangled states and maximally entangled measurements.}
\end{figure}

In the supplementary material, we list the waveplate angles used to target each of the 60 two-qubit stabilizer states and each of the 60 stabilizer effects in our experimental setup. We also highlight the subsets of these that correspond to targeting elements of the {\em product sector}, denoted $\mathcal{S}_{\rm prod}$ and $\mathcal{E}_{\rm prod}$, as well as those that lie in the {\em real-amplitude sector}, denoted $\mathcal{S}^{\rm real}$ and $\mathcal{E}^{\rm real}$. The intersection of these subsets (those that are both product and real) are denoted $\mathcal{S}^{\text{real}}_{\text{prod}}$ and $\mathcal{E}^{\text{real}}_{\text{prod}}$ respectively.

\subsection{The tensor product structure}\label{sec:TPS}



%
%

 In some experiments, it may be that the tensor product structure is encoded in the preparation procedures.  For a pair of systems, $A$ and $B$, there may be the possibility of implementing 
 independent procedures on each subsystem, so that the state on $AB$ includes no correlations between $A$ and $B$.  In this case, one is justified in constraining the GPT state vectors on $AB$ in the fit 
 to those that are strictly factorizing,  i.e., to $\bm{s} \in V$ such that there exists a vector $\bm{s}_A \in V_A$ and a vector $\bm{s}_B \in V_B$ such that $\bm{s}= \bm{s}_A \otimes \bm{s}_B$ 

The experiment we conducted, however, is not of this type. 
Although we are targeting the preparation of two-photon polarization product states,  
small differences between the ideal experimental set-up and what is actually realized in the experiment can cause the realized state to  {\em not} be precisely of product form. 
This can occur, for instance, if the polarization of the pump laser deviates from the one we are targeting 
or if the relative orientation of the two nonlinear crystals deviates from perfect orthogonality.
 Similarly, although we are targeting product effects, what is realized in practice is highly unlikely to be precisely of product form. This is often due to imperfect two-photon interference, which can result from non-ideal extinction ratio in the polarizing beam splitter used to bring the photons together, or small angular misalignments in the optical components.
 
 Because of these considerations, we do not look to the preparation or measurement procedures in our experiment to find the empirical counterpart of the tensor product structure.
Rather, we define the tensor product structure in terms of the {\em transformation procedures}. This is because in our experiment, transformations are applied to each photonic mode independently, such that the effective transformation on the pair of photonic modes is expected to be factorizing across the modes. More precisely, although there are still some exotic possibilities for a deviation from perfect factorization, the magnitude of such deviations are expected to be much smaller than those that arise in the case of the preparations or the measurements.  Thus, although it is still an approximation to assume factorization for the transformations, it is a much better approximation than to do so for the preparations or measurements. 

One can decompose the preparation stage of the experiment into an initial preparation of the composite system followed by a transformation thereon.  Similarly, one can decompose the measurement stage of the experiment into a measurement of the composite system that is {\em preceded}  by a transformation thereon.  
It is these transformation procedures that we use  
to identify the tensor product structure.
A transformation on the polarization degree of freedom of a photonic mode is achieved by a sequence of quarter and half waveplates in that mode, with variations in the orientation of these waveplates achieving variations in the transformation that is implemented.  For each mode, the angles of the waveplates in that mode determine the local transformation applied there (independently of what is done to the other mode).
 We therefore model the overall bipartite transformations 
by {\em factorizing} GPT matrices, i.e., by matrices $T: V\rightarrow V $ of the form $T= T_A \otimes T_B$
for some $T_A : V_A \rightarrow V_A$ and some $T_B : V_B \rightarrow V_B$.

In summary, when we do our fit via GPT tomography, we impose the constraint that the GPT matrices representing the transformation procedures we implemented are of product form, but we do not impose such a constraint on the preparations and measurements. More precisely,  
for the preparation and measurement procedures where an ideal implementation {\em would be} factorizing (i.e., those where we were targeting product quantum states and product quantum effects), we expect experimental deviations from the ideal and so we do {\em not} impose the constraint that the GPT states and effects representing these procedures should be of product form. Nonetheless, tomographic locality is based on the distinction between product and entangled states, thus the next step of our analysis is to characterize a set of GPT state vectors and GPT effect vectors that {\em are} perfectly factorizing.
These do not represent any of the preparations or measurements we realized experimentally but are rather inferred to exist by virtue of being included among the {\em probabilistic mixtures}  of those we realized experimentally. 
In other words, we characterize a set of product GPT state vectors and product GPT effect vectors 
\blk 
using the technique of {\em secondary procedures} introduced in Ref.~\cite{mazurek2021experimentally}.

\subsection{Finding secondary state and effect vectors that are strictly factorizing}\label{sec:secondary}

As discussed earlier, due to imperfections in our experimental setup, preparations targeting product states on $AB$ (i.e., an element of $\mathcal{S}_{\rm prod}$)
likely introduced a small amount of correlation between $A$ and $B$, and hence realized a state that was not strictly factorizng.  
(Note that the distinction between factorizing and nonfactorizing is evaluated relative to the tensor product structure that emerges from the fit under the assumption that the transformations are precisely factorizing.) We denote by ${\mathcal{S}}$ the set of GPT state vectors  that represent (within the best-fit model) the full set of preparation procedures realized in the experiment, and by ${\mathcal{S}}_{\rm prod} \subset \mathcal{S}$ the subset corresponding to preparations targeting product states. In standard quantum theory, this set corresponds to 36 out of the 60 total preparations; in the real-amplitude sector, it corresponds to 16 out of 24. As noted above, none of the elements of $\mathcal{S}_{\rm prod}$ are expected to be strictly factorizing. Nonetheless, such factorizations can exist within the convex hull of the full set $\mathcal{S}$. Therefore, for each $\bm{s} \in \mathcal{S}_{\rm prod}$, we seek the closest GPT state within the convex hull of $\mathcal{S}$ that is strictly factorizing, where the measure of closeness is measured using the 2-norm in the GPT vector space. We refer to this state as the secondary GPT state vector, denoted $\bm{s}^{\,\rm sec}$, and the full set of such vectors is denoted $\tilde{\mathcal{S}}_{\rm prod}$.
	 
Thus, we implement the following algorithm to determine $\bm{s}^{\rm sec}$ for each of the elements of $\mathcal{S}_{\rm prod}$:
		\begin{equation}
        \label{secproc}
		\begin{aligned}
		&\forall \bm{s} \in \mathcal{S}_{\rm prod}\\
			\textrm{MIN:} \quad & |\bm{s}^{\,\rm sec}-\bm{s}|_2 \\
			\textrm{Subject to:}
			 \quad & \bm{s}^{\,\rm sec} 
			 = \sum_{\bm{s} \in \mathcal{S}} w_{\bm{s}} \bm{s}\\
             & \sum_{\bm{s} \in \mathcal{S}} w_{\bm{s}} = 1\\
			& \forall \bm{s} \in \mathcal{S} :  0 \leq w_{\bm{s}} \leq 1\\
			&\exists \bm{s}_A, \bm{s}_B : \bm{s}^{\,\rm sec} = \bm{s}_A \otimes \bm{s}_B\\
		\end{aligned}
	\end{equation}
where $|\bm{v}|_2$ denotes the 2-norm of vector $\bm{v}$, i.e,  $|\bm{v}|_2= \sum_{\alpha} v_{\alpha}^2$. Note that the GPT states $\{\bm{s}_i\}_{i=1}^{n}$ that we allow in our convex decomposition of $\bm{s}^{\rm sec}$ are drawn from  
the full set $\mathcal{S}$ of experimentally realized GPT state vectors, rather than just those in $\mathcal{S}_{\rm prod}$. This ensures that we are probing over the full convex hull
of physically realized 
states when constructing each secondary state.

An analogous algorithm is applied to find, for each $\bm{e} \in \mathcal{E}_{\rm prod}$, a secondary GPT effect vector that is strictly factorizing and which we denote by $\bm{e}^{\,\rm sec}$. The set of secondary GPT effect vectors will be denoted by 
	$\tilde{\mathcal{E}}_{\rm prod}$.



 We implement the secondary procedures technique also when we do the data analysis for preparations and measurement that target the real-amplitude sector. The corresponding algorithm for the preparations, for instance, is simply the one described in Eq.~\eqref{secproc} but with $\mathcal{S}^{\rm real}$ substituted for $\mathcal{S}$ and $\mathcal{S}^{\rm real}_{\rm prod}$ substituted for $\mathcal{S}_{\rm prod}.$ We refer to the resulting sets of product states and effects as $\tilde{\mathcal{S}}^{\text{real}}_{\text{prod}}$ and $\tilde{\mathcal{E}}^{\text{real}}_{\text{prod}}$ respectively.  

\subsection{
Dimension of the embedding vector space and effective ranks}

Tomographic locality fails in a theory if there is a deficit in the dimension of the span of the product GPT state vectors compared to the dimension of the span of all the GPT state vectors. Experimentally, therefore, we must estimate this pair of dimensions. 

Recall that $\mathcal{S}$ is the set of all GPT state vectors for preparations realized in the experiment. Our analysis makes the assumption that $\mathcal{S}$ is a sufficiently diverse 
set that it spans the same vector space as the full set of states allowed by the theory governing our experiment. The dimension of the span of the vectors in $\mathcal{S}$ necessarily coincides with the rank of the matrix $D^M$ for the best-fit model $M$. 


Our analysis also makes the assumption that $\tilde{\mathcal{S}}_{\rm prod}$ is a sufficiently generic set that it spans the same vector space as would the full set of product states allowed by the theory governing our experiment.

In the case of the real-amplitude sector, the dimension of the span of the GPT state vectors in $\mathcal{S}^{\rm real}$ can be (and is expected to be according to quantum theory) less than the span of the GPT state vectors in $\mathcal{S}$. Similarly with  $\tilde{\mathcal{S}}_{\rm prod}^{\rm real}$ relative to $\tilde{\mathcal{S}}_{\rm prod}$.  



Given that we presume that the system being probed is bipartite, 
the GPT representation for the bipartite system must embed into a vector space that is a {\em tensor product} of the vector spaces that embed the GPT representations for each component system.  Moreover, we assume that the subsystems are of the same type, so that they have the same dimension of embedding vector space, say $d$, and consequently the embedding vector space for the composite has dimension $d^2$. 


Whenever it happens that the GPT state vectors representing some {\em idealized} set 
span a strict subspace of the full GPT vector space,
the GPT state vectors that represent what are physically realized in the experiment will only {\em approximate} this span.  That is, we do not expect the linear dependence relations among the latter to reproduce exactly those of the former. 

Consequently, the dimension of span that one computes directly should be replaced by an estimate of {\em effective dimension} that 
takes into account
small deviations from linear dependence.  We now describe how we make this estimate.

First of all, we note that in order to estimate the effective dimension, one can equally well look at sets of GPT effect vectors rather than sets of GPT state vectors.  
The dimension of the span of all GPT state vectors $\mathcal{S}$
is equivalent to that of the span of all GPT effect vectors $\mathcal{E}$
simply because these sets 
have been obtained 
 as the rows of a matrix $S$ and the columns of a matrix $E$ in a decomposition $D^{M}=SE$ where  $D^{M}$ is the matrix of probabilities predicted by the model. 
In the case of product states and effects, however, where we allow for the possibility of a dimensional discrepancy with the full sets, or for the case of the real-amplitude quantum states and effects (as well as the case of those that are both real-amplitude and product), where we also allow (and expect) such a discrepancy, it is conceivable that for a given threshold for estimating the effective dimension, one might in principle get different results using just the states or just the effects. This is because the spans of the sets of effects 
can have slightly different deviations from precise linear dependence relations than those of the states.  

We therefore 
use {\em both} the sets of state vectors and the sets of effect vectors {\em together} to estimate the effective dimension. We do so as follows. 
 Define a 2-place function $D$ that takes as its inputs a set of GPT state vectors $\{\bm{s}_i\}_{i=1}^{n}$ and a set of GPT effect vectors $\{\bm{e}_j\}_{j=1}^{n}$ and returns the matrix of probabilities obtained by contraction of these, i.e., the matrix 
  $\mathsf{D}(\{\bm{s}_i\}_{i=1}^{n},\{\bm{e}_j\}_{j=1}^{n})$ has $(i,j)$th component equal to 
 ${\bf s}_i \cdot {\bf e}_j$.

Then, for each of the matrices $\mathsf{D}(\tilde{\mathcal{S}}_{\text{prod}},\tilde{\mathcal{E}}_{\text{prod}})$, $\mathsf{D}(\mathcal{S}^{\rm real},\mathcal{E}^{\rm real})$ and $\mathsf{D}(\tilde{\mathcal{S}}^{\rm real}_{\text{prod}},\tilde{\mathcal{E}}^{\rm real}_{\text{prod}})$, we compute the singular value decomposition.  To define  the {\em effective rank} of each such matrix, we set a threshold below which the singular values are taken to be effectively zero (to be discussed further on), and count the number of singular values that are judged to be {\em not} effectively zero.

Finally, comparison of the rank of $\mathsf{D}(\mathcal{S},\mathcal{E})$ (equivalently, the rank of the best-fit matrix $D^M$) with the effective rank of $\mathsf{D}(\tilde{\mathcal{S}}_{\text{prod}},\tilde{\mathcal{E}}_{\text{prod}})$ yields the verdict concerning whether tomographic locality holds in our experiment.  Similarly, comparison of the effective rank of  $\mathsf{D}(\mathcal{S}^{\rm real},\mathcal{E}^{\rm real})$ with that of $\mathsf{D}(\tilde{\mathcal{S}}^{\rm real}_{\text{prod}},\tilde{\mathcal{E}}^{\rm real}_{\text{prod}})$ yields the verdict concerning whether tomographic locality holds in our simulation of the real-amplitude sector.

\blk

\subsection{Summary of the steps in the data analysis}

The flowchart in Fig.~\ref{fig:four_subfigures} provides a broad outline of the steps in the data analysis procedure. It begins with applying GPT tomography in order to reconstruct the set $\mathcal{S}$ of GPT state vectors  and the set $\mathcal{E}$ of GPT effect vectors in the best-fit GPT model of the experimental data (with the rank of the model being determined by a train-and-test methodology).

From these, we extract the subset of states and effects $\mathcal{S}_{\text{prod}}$ and $\mathcal{E}_{\text{prod}}$, corresponding to the preparations and measurements wherein we were targeting product states and product effects. These subsets serve to define the sets of GPT state vectors and GPT effect vectors that are strictly factorizing, using the technique of secondary procedures. We refer to these sets as $\tilde{\mathcal{S}}_{\text{prod}}$ and $\tilde{\mathcal{E}}_{\text{prod}}$. We can then construct the matrix $\mathsf{D}(\tilde{\mathcal{S}}_{\text{prod}}, \tilde{\mathcal{E}}_{\text{prod}})$ (by taking the inner products of each state vector with each effect vector). Finally, we compute the effective rank of this matrix. 


If these matrices are found to have the same rank, then tomographic locality holds. If not, this signals the failure of tomographic locality in this regime.

\begin{figure}[ht!]
    \centering
	\includegraphics[width=1\linewidth]{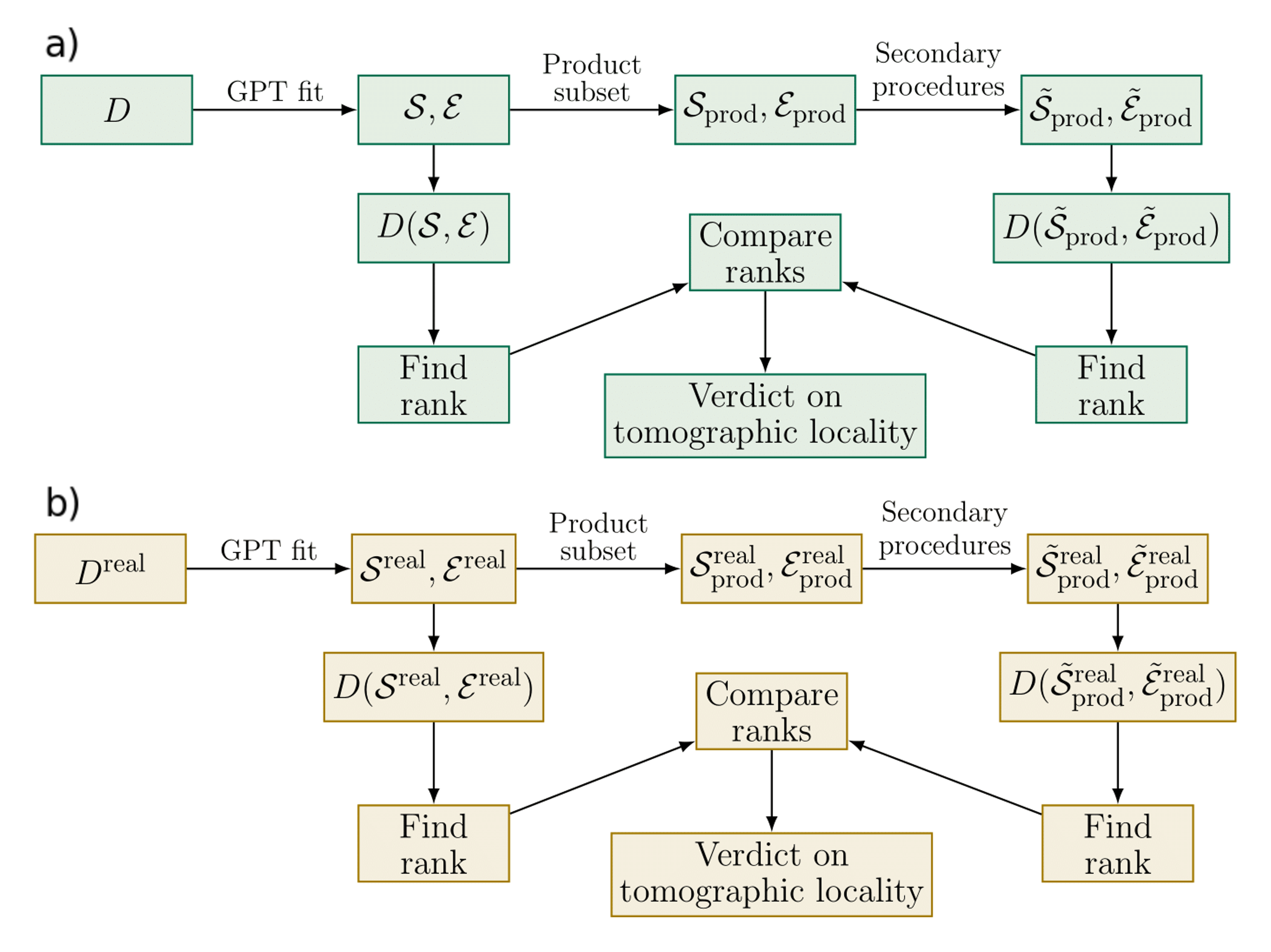}
    \caption{\justifying Flowcharts summarizing the procedures for testing tomographic locality in a bipartite GPT framework. a) Procedure to determine whether or not tomographic locality holds for a bipartite photonic system. From the raw experimental data $D$, we use a train-and-test GPT fitting procedure to extract $\mathcal{S}$ and $ \mathcal{E}$. From these, we extract the subsets $\mathcal{S}_{\text{prod}}$ and $ \mathcal{E}_{\text{prod}}$, corresponding to the targeted product sets. We then apply secondary procedures to obtain strictly factorizing approximations, $\tilde{\mathcal{S}}_{\text{prod}}$ and $ \tilde{\mathcal{E}}_{\text{prod}}$. The resulting ranks of $\mathsf{D}({\mathcal{S}}, {\mathcal{E}})$ and $\mathsf{D}(\tilde{\mathcal{S}}_{\text{prod}}, \tilde{\mathcal{E}}_{\text{prod}})$ are then compared to assess whether tomographic locality is upheld. b) Analogous procedure but applied to the raw experimental data  $D^\text{real}$ obtained by targeting states and effects that are confined to the real-amplitude sector.}
    \label{fig:four_subfigures}
\end{figure}

\subsection{Expectations}

If quantum  theory is the correct  theory of nature, then the polarization degree of freedom of a single photon in a well-defined mode constitutes a qubit and is thus described by a 4-dimensional real vector space in the GPT formalism:  this is the 4-dimensional vector space of Hermitian operators on a qubit, spanned by the Pauli operators $I$, $X$, $Y$ and $Z$. The polarization degree of freedom of a single photon mode containing exactly one photon is therefore expected to be described in the GPT formalism by a 4-dimensional vector space, equivalent to the Bloch sphere representation. For a pair of modes, it is the tensor product of two such spaces (the space of Hermitian operators on two qubits), hence 16-dimensional.

 To allow for the possibility that the experiment reveals the need for a lower or higher dimensionality of the embedding vector spaces (compared to what is expected according to quantum theory), we consider a slate of models where the embedding vector space dimension for each subsystem ranges over $d \in \{1,2,3,4,5\}$, and consequently the embedding vector space dimension of the pair of modes ranges  over $d^2 \in \{1,4,9,16,25\}$.

If the quantum predictions are upheld, then we expect that when we perform the optimization over possible dimensions of the embedding vector space and fit the $60 \times 60$ matrix of relative frequencies $D$ to a best-fit model $D^M$ of each dimension, the most predictive such model in the train-and-test methodology has $d^2 =16$. 
That is, we expect that the GPT vector space embedding the set of GPT state vectors of the best-fit model, $\mathcal{S}$, and the set of GPT effect vectors of the best-fit model, 
$\mathcal{E}$, has dimension 16, and consequently that the matrix of probabilities $\mathsf{D}(\mathcal{S},\mathcal{E})$ has full rank in the embedding vector space, i.e., rank $16$. 


Because the product stabilizer states in quantum theory also span the full space of Hermitian operators on two qubits, we also expect 
the effective rank of
 the $36 \times 36$ matrix $\mathsf{D}(\tilde{\mathcal{S}}_{\text{prod}},\tilde{\mathcal{E}}_{\text{prod}})$ of probabilities inferred for the strictly factorizing GPT state and effect vectors to be $16$. To see evidence for 
 a failure of tomographic locality, the data analysis must find that, contrary to quantum expectations, the ranks of $\mathsf{D}(\mathcal{S},\mathcal{E})$ and $\mathsf{D}(\tilde{\mathcal{S}}_{\text{prod}},\tilde{\mathcal{E}}_{\text{prod}})$ are different.
 
 

  



Now consider the case where  we restrict attention to preparations and measurements targeting the real-amplitude sector of the stabilizer formalism, thereby simulating what would occur if the correct theory of nature were real-amplitude quantum theory. 

The product stabilizer states in the real-amplitude sector span a 9-dimensional subspace of the space of Hermitian operators on two qubits, namely, the space spanned by products of Pauli operators when these are restricted to the set $\{ I,X,Z\}$ on each qubit. 
A pair of qubits restricted to the real-amplitude sector is described in the GPT formalism  by a real vector space that has dimension 10. This is because the span of two-qubit Hermitian operators that are real-amplitude includes a tensor product of single-qubit Hermitian operators that are {\em not} themselves real-amplitude, namely, the product of the Pauli $Y$ operator with itself, $Y \otimes Y$. 
For instance, the pair of stabilizer states $\rho_{\pm} := \tfrac{1}{4} (I \otimes I \pm Y \otimes Y)$ are in the real-amplitude sector and have nonzero components on these Hermitian operator. \footnote{It is worth noting that because the measurement of Pauli $Y$ one one mode is not in the real-amplitude sector of the local measurements, this pair of states cannot be distinguished by product effects in the real-amplitude sector, and this indistinguishability is an operational signature of the failure of tomographic locality. }

As noted above, we perform the same analysis as for the full stabilizer set but this time starting with the $24 \times 24$ matrix of relative frequencies $D^{\rm real}$ (i.e., the submatrix of $D$ where we targeted the real-amplitude states and effects). 

When we perform a optimization over possible dimensions of the embedding vector space and find the best-fit models for each such dimension, we again expect our train-and-test methodology to yield $d^2=16$ because this is the smallest square dimension that can contain the 10-dimensional space spanned by the real-amplitude stabilizer states and effects. 
That is, we expect the GPT state vectors in $\mathcal{S}^{\text{real}}$ and the GPT effect vectors in $\mathcal{E}^{\text{real}}$ to again be 16-dimensional.  The matrix of probabilities $\mathsf{D}(\mathcal{S}^{\text{real}},\mathcal{E}^{\text{real}})$ that these define, however, is expected to have an effective rank of $10$. 

Furthermore, because the real-amplitude stabilizer states that are of product form span a 9-dimensional subspace of this 10-dimensional space, we expect that after applying the secondary procedures technique to infer sets of strictly factorizing states and effects,  $\tilde{\mathcal{S}}^{\text{real}}_{\rm prod}$ and $\tilde{\mathcal{E}}^{\text{real}}_{\rm prod}$, the $16 \times 16$ matrix of probabilities that these define, $\mathsf{D}(\tilde{\mathcal{S}}^{\text{real}}_{\rm prod},\tilde{\mathcal{E}}^{\text{real}}_{\rm prod})$, will  have an effective rank of 9.

In summary, when we do the data analysis while targeting the stabilizer states and effects to the real-amplitude sector, the signature of the failure of tomographic locality we expect to see is a gap in the effective rank of
$\mathsf{D}({\mathcal{S}}^{\text{real}},{\mathcal{E}}^{\text{real}})$ 
and the effective rank of 
 $\mathsf{D}(\tilde{\mathcal{S}}^{\text{real}}_{\text{prod}},\tilde{\mathcal{E}}^{\text{real}}_{\text{prod}})$.

\blk 

\blk


\subsection{Reducing the number of parameters in the model}\label{4f}

One final nuance regarding the data analysis must be considered before we turn to describing the results: how to parameterize the GPT states and GPT effects of a given model.


The preparation stage of the experiment begins by targeting one of a pair of two-qubit states with $\rm{WP}_1$, namely, $|H\rangle|H\rangle$ or $\frac{1}{\sqrt2}(|H\rangle|H\rangle+|V\rangle|V\rangle)$. This is achieved by adjusting $\rm{WP}_1$ to either \( 0^\circ \) or \( 22.5^\circ \), toggling between the two preparations. This is then followed by two waveplates on photonic mode A, denoted $\rm{WP}_2$ and $\rm{WP}_3$, and three waveplates on photonic mode B, denoted $\rm{WP}_4$, $\rm{WP}_5$, $\rm{WP}_6$. The set of possible angles to which each waveplate $\rm{WP}_{\mu}$ can be set will be denoted $S_{\mu}$. We have $S_1 := \{0,22.5\}$, $S_{2} := \{-22.5,0,22.5,45\}$, $S_3 := \{0,45\}$, $S_{4} := \{-22.5,0,22.5\}$, $S_5 := \{-45,45\}$, and $S_{6} := \{0,22.5\}$.

We can therefore model the preparation stage in the GPT model as follows. The initial preparation is modeled in the GPT by one of a pair of GPT state vectors, denoted $\bm{s}^{\rm init}_{\phi_1} \in V$ where ${\phi_1}\in\{0,1\}$ (here, ${\phi_1}=0$ corresponds to the targeting $|H\rangle|H\rangle$ and ${\phi_1}=1$ corresponds to targeting $\frac{1}{\sqrt2}(|H\rangle|H\rangle + |V\rangle|V\rangle$)). Each waveplate in mode $A$ is modeled in the GPT by a matrix acting on the GPT vector space $V_A$. The matrix associated to the transformation induced by $\rm{WP}_{\mu}$ being set at angle $\phi_{\mu}$ (for $\mu \in \{2,\dots,11\}$) is denoted $T^{(\mu)}_{\phi_{\mu}}$. 

 We can consequently associate to  the index $i\in \{1,\dots, 60\}$ (labeling the preparations) a set of indices $(\phi_1,\phi_2,\phi_3,\phi_4, \phi_5,\phi_6)$, and the overall GPT state vector $\bm{s}_{i}$ can be expressed as 
 $$\bm{s}_{i}= \left[ ( T_{\phi_3}^{(3)} \circ T_{\phi_2}^{(2)}) \otimes ( T_{\phi_6}^{(6)}\circ T_{\phi_5}^{(5)} \circ T_{\phi_4}^{(4)})  \right] \circ \bm{s}^{\,\rm init}_{\phi_1}.$$
To parametrize the 60 GPT state vectors,  therefore, it suffices to use the parameters $
{\bf par} = \{ \bm{s}^{\,\rm init}_{\phi_1}:{\phi_1} \in \{1,2\} \} \cup \left( \bigcup_{\mu=2}^{6}\{ T_{\phi_{\mu}}^{({\mu})}:\phi_{\mu} \in S_{\mu} \blk \} \right)$.
We write $\bm{s}_i({\bf par})$ to emphasize that each $\bm{s}_i$ can be defined in terms of a choice of ${\bf par}$.


Next, we turn to the measurement stage of the experiment. At the end of this stage, one of a pair of two-qubit effects are realized, namely, the projector onto $|H\rangle|H\rangle$ or the projector onto $\frac{1}{\sqrt2}(|H\rangle|H\rangle+|V\rangle|V\rangle)$. This is achieved by adjusting $WP_{12}$ to either \( 0^\circ \) or \( 22.5^\circ \), toggling between the two projectors.
This is preceded by two waveplates on photonic mode A, denoted $\rm{WP}_7$ and $\rm{WP}_{8}$, and three waveplates on photonic mode B, denoted $\rm{WP}_9$, $\rm{WP}_{10}$, and $\rm{WP}_{11}$. The set of possible angles to which each of these waveplates can be set to are, $S_7 := \{0,45\}$, $S_{8} := \{-22.5,0,22.5,45\}$, $S_9 := \{0,22.5\}$, $S_{10} := \{-45,45\}$, $S_{11} := \{-22.5,0,22.5\}$, $S_{12} := \{0,22.5\}$. 


It follows that we can associate to  the index $j\in \{1,\dots, 60\}$ (labeling the measurements) a set of indices $(\phi_7,\phi_8, \phi_9,\phi_{10},\phi_{11},\phi_{12})$, and the overall GPT effect vector $\bm{e}_{j}$ can be expressed as 
 $$\bm{e}_{j}= \bm{e}^{\,\rm fin}_{\phi_{12}} \circ \left[ ( T_{\phi_7}^{(7)} \circ T_{\phi_8}^{(8)} ) \otimes (T_{\phi_{9}}^{(9)} \circ T_{\phi_{10}}^{(10)} \circ T_{\phi_{11}}^{(11)})  \right].$$
 To parametrize the 60 GPT effect vectors, it suffices to use the parameters $
{\bf par}' = \{ \bm{e}^{\,\rm fin}_{\phi_{12}}:{\phi_{12}} \in \{1,2\} \} \cup \left( \bigcup_{\mu=7}^{11}\{ T_{\phi_{\mu}}^{({\mu})}:\phi_{\mu} \in S_{\mu} \blk \} \right)$.
We write $\bm{e}_j({\bf par}')$ to emphasize that each $\bm{e}_j$ can be defined in terms of a choice of ${\bf par}'$. 

A given model $M$ with GPT dimension $k$,
parameter values ${\bf par}$ for the preparation stage, and parameter values ${\bf par}'$ for the measurement stage 
 defines a rank-$k$ matrix of probabilities $D^M$ via 
  $$D^M_{ij} =   \bm{s}_i({\bf par})\circ \bm{e}_{j}({\bf par}').$$ 
When implementing an optimization over models $M$ with a given GPT dimension, therefore, it suffices to vary over the possible choices of ${\bf par}$ and ${\bf par}'$.

This parameterization is significantly more economical than a brute-force approach that would assign independent transformation matrices to each of the GPT state and effect vectors. By instead modeling these in terms of a smaller set of mutual transformation matrices, we drastically reduce the number of parameters needed. For a 16 dimensional GPT, the brute-force approach requires $3902$ parameters whereas our structured parameterization requires only $478$. \footnote{In a given method, the total number of parameters is calculated as the sum of those required to specify the states, the transformations acting on the states and measurements, and the measurements themselves. For the brute force method, this becomes: 2($d^2-1$) + $2 d^2\times 60$ + $2d^2 \times 60$ + $2d^2$. For the structured parameterization, it is: 2($d^2-1$) + $13d^2$ + $13d^2$ + $2d^2$. In both methods, we assume that every transformation is a tensor product of a pair of $d \times d$ matrices.}


\blk


\section{Results}

In Fig.~\ref{fig:freqMatrix}(a), we plot the relative frequency of the positive measurement outcome for each of the $3600$ preparation-measurement pairings, i.e., the components of the $60\times 60$ raw data matrix $D$. 



\begin{figure}[h!]
    \centering
	\includegraphics[width=1\linewidth]{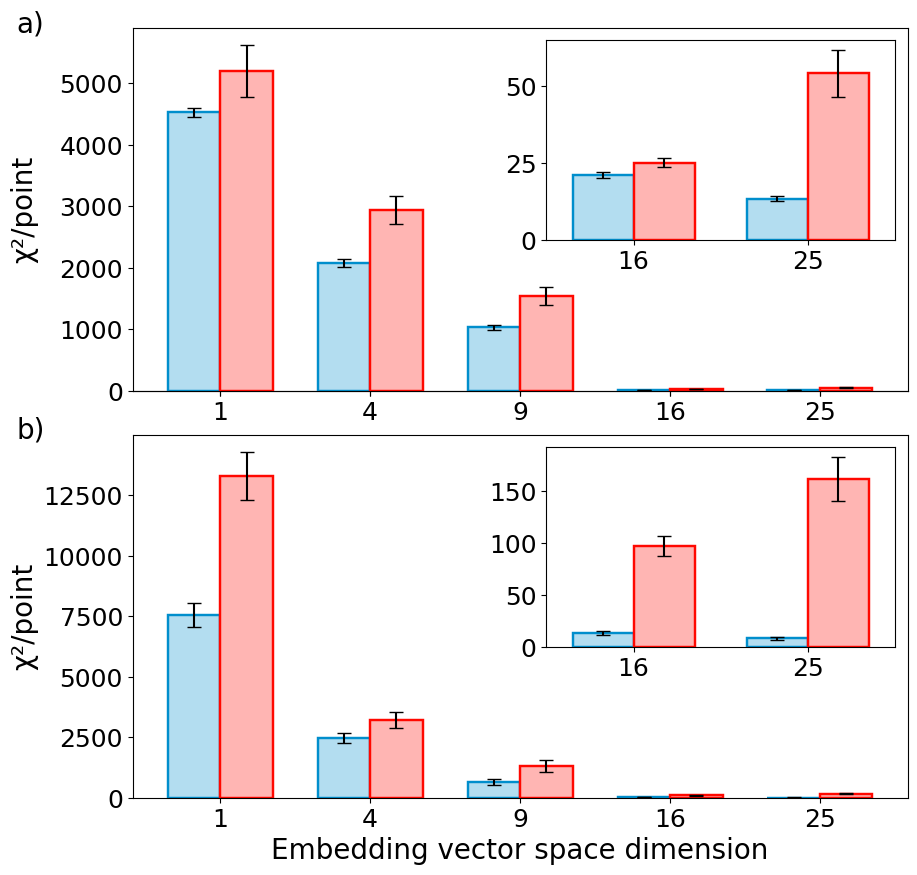}
	
	\caption{\justifying 
    \label{fig:traintest}
	  Training and test errors of the best-fit GPT models of different dimensions of the embedding GPT vector space (i.e., $d^2 \in \{1,4,9,16,25\}$) for a) $D$ and b) $D^{\text{real}}$. Blue bars represent the training error, $(\chi^2)^{\rm train}$, while red bars represent the test error, $(\chi^2)^{\rm test}$.  In both cases, the minimum test error is achieved for the embedding vector space dimension $d^2 = 16$, consistent with the quantum expectation.  The error bars represent one standard deviation in the mean.}
\end{figure}

\begin{figure*}[ht!]
    \centering
	\includegraphics[width=1\linewidth]{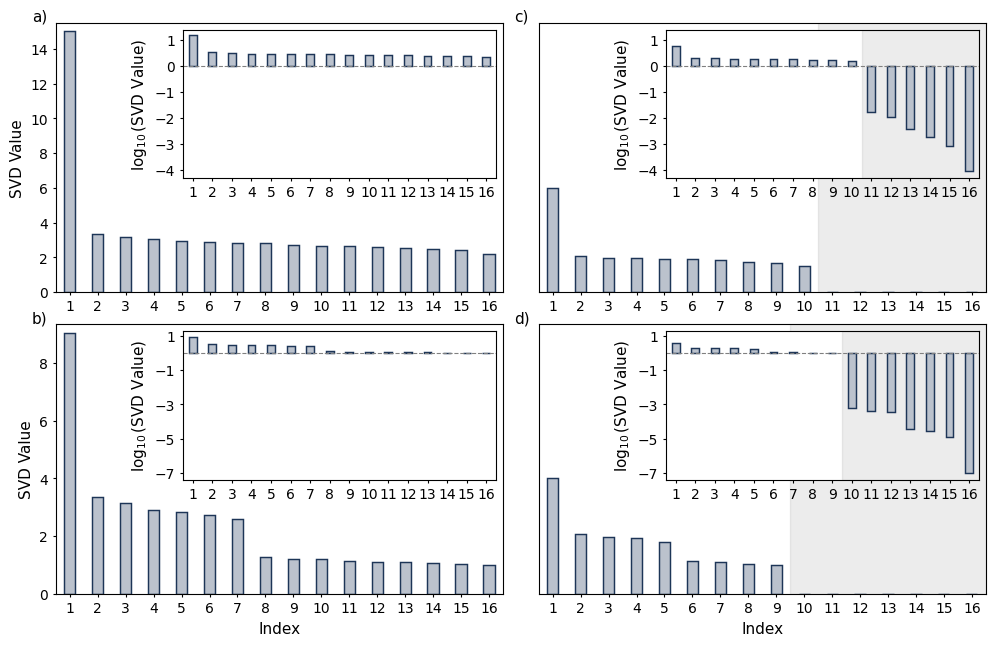}

	   \caption{\justifying Estimating the effective ranks. The different panels plot the set of singular values of a) $\mathsf{D}(\mathcal{S},\mathcal{E})$ b) $\mathsf{D}(\tilde{\mathcal{S}}_{\text{prod}},\tilde{\mathcal{E}}_{\text{prod}})$ c) $\mathsf{D}(\mathcal{S}^{\text{real}},\mathcal{E}^{\text{real}})$ d) $\mathsf{D}(\tilde{\mathcal{S}}^{\text{real}}_{\text{prod}},\tilde{\mathcal{E}}^{\text{real}}_{\text{prod}})$. The shaded regions denote singular values that are below $10^{-1.6}$, which we treat as effectively zero based on empirical estimates of statistical noise in the experiment. The agreement in rank between a) and b) supports the validity of tomographic locality in quantum theory. The dimensional mismatch between c) and d) confirms the expectation that data simulating real-amplitude quantum theory exhibits a failure of tomographic locality. 
}
       \label{fig:four_subfigures}
\end{figure*}

In Fig.~\ref{fig:traintest}(a), we plot the training and test errors for the best-fit GPT models to $D$ for each of  
the possible dimensions of the embedding vector space used by the model.

Each model was fit by minimizing the training error based on choosing the training set to be a subset consisting of 90\% of the components of the raw data matrix, chosen uniformly at random, with the remaining 10\% making up the test set. 
This cross-validation procedure was repeated 10 times, each with a different random sampling.
The values plotted in the figure represent the average over these 10 trials, with the error bars corresponding to the standard deviation of them. 

We see that, consistent with the quantum expectation, an embedding vector space dimension of $d^2=16$ 
 minimizes the test error and hence is the most predictive of the models we considered.
We also find that, again as one expects under the assumption of the correctness of quantum theory,
the only choices of dimension that can achieve a low training error are  $d^2 \ge 16$, i.e., $d^2 \in \{16,25\}$, while for $d^2 \le 9$ the training error is very large and indicates underfitting of the data by the models.
Comparing $d^2 =25$ with $d^2 =16$, we see that the test error is smaller for $d^2=16$, indicating that it has better predictive power than the model with $d^2 =25$.  Furthermore, the fact that the model with $d^2 =25$ has a {\em lower} training error than the one with $d^2=16$ suggests that the former {\em overfits} the data relative to the latter, and that it is this overfitting that leads to its worse predictive power.


To see why one might expect overfitting of the $d^2 =25$ 
model relative to the $d^2 =16$
model, it is worth noting that the number of parameters in the model grows significantly with every unit increase in $d$. For instance, if we are considering all states, so that $m = n = 60$, the number of parameters in the $d^2 =25$ 
model (748 parameters) greatly exceeds the number in the $d^2 =16$ 
model (478 parameters), as determined by the parameterization of the experimental procedures described in 
Sec.~\ref{4f}.


Consequently, in our evaluation of the evidence for tomographic locality, we will use the best-fit model with $d^2 =16$.



In  Figs.~\ref{fig:four_subfigures}(a) and (b), we plot the singular values of the matrices $\mathsf{D}(\mathcal{S},\mathcal{E})$ and
$\mathsf{D}(\tilde{\mathcal{S}}_{\text{prod}},\tilde{\mathcal{E}}_{\text{prod}})$ in descending order.
All the singular values of both of these matrices are order 1 or larger, i.e., none are effectively zero, thereby confirming the expectation that  both have an
effective rank of 16. 
The lack of any discrepancy between these effective ranks means that our experiment does not provide any evidence for a failure of tomographic locality. 

We turn now to the experiment wherein we simulated what would be observed if real-amplitude quantum theory were the correct theory of nature (as described in Sec.~\ref{3a}).    In Fig.~\ref{fig:freqMatrix}(b), we plot the $24 \times 24$ submatrix of $D$ describing the relative frequency of the positive measurement outcomes for each preparation-measurement pairing wherein we targeted preparations and measurements in the  real-amplitude sector.  These are the components of the matrix $D^{\rm real}$. 
In Fig.~\ref{fig:traintest}(b), we plot the training and test errors for the best-fit GPT models to $D^{\rm real}$ for each of  the possible dimensions of the embedding vector space used by the model, following the same train-and-test methodology that was applied to $D$. 

Again, we see that the model that has the most predictive power is the one with $d^2 =16$, so that this is the one that defines $\mathcal{S}^{\rm real}$ and $\mathcal{E}^{\rm real}$.  From these, we determine $\tilde{\mathcal{S}}^{\rm real}_{\rm prod}$ and $\tilde{\mathcal{E}}^{\rm real}_{\rm prod}$.  We can then compute $\mathsf{D}(\mathcal{S},\mathcal{E})$ and
$\mathsf{D}(\tilde{\mathcal{S}}_{\text{prod}},\tilde{\mathcal{E}}_{\text{prod}})$.  


In  Figs.~\ref{fig:four_subfigures}(c) and (d), we plot the singular values of these matrices in descending order.
For $\mathsf{D}({\mathcal{S}}^{\text{real}},{\mathcal{E}}^{\text{real}})$, we see that the 10 largest singular values are order 1 or larger, and the rest have a value of $10^{-1.6}$ or smaller. 
 Meanwhile, for $\mathsf{D}(\tilde{\mathcal{S}}^{\text{real}}_{\text{prod}},\tilde{\mathcal{E}}^{\text{real}}_{\text{prod}})$, it is only the 9 largest singular values that have order 1 or larger, while the rest have a value of $10^{-3.1}$ or smaller.  
 If we take $10^{-1}$ to be the threshold below which a singular value is no longer counted as contributing to the effective rank, 
 (more on this choice below), then we conclude that the effective rank of $\mathsf{D}({\mathcal{S}}^{\text{real}},{\mathcal{E}}^{\text{real}})$ is 10, while that of $\mathsf{D}(\tilde{\mathcal{S}}^{\text{real}}_{\text{prod}},\tilde{\mathcal{E}}^{\text{real}}_{\text{prod}})$ is 9.  
 
Consequently, we find a mismatch in our estimate of the dimension of the span of the full set of real-amplitude states (effects) and the dimension of the span of the subset of real-amplitude states (effects) that are of product form, which is the signature of the failure of tomographic locality. 
We thereby confirm the expectation that a simulation of real quantum theory should exhibit such a failure.  

Singular values smaller than $10^{-1}$ are not counted towards the effective rank, with this threshold being justified by an empirical estimate on the statistical noise in our photon count data. To obtain this estimate, we used the fact that each row of our data matrix $D$ corresponds to an overcomplete set of 36 tomographic measurements in the product sector, where our state preparation and measurement fidelities were highest. For each row, we performed tomography using 35 of the 36 measurements to predict the expected photon count of the omitted measurement. We then compared this prediction to the actual measurement outcome, with the deviation providing an estimate of the statistical noise. This procedure was repeated across all states, each time leaving out a different randomly selected measurement. The results indicated that the statistical noise level in our data is approximately five times larger than what would be expected from Poissonian statistics alone. We simulated theoretical data with noise at this level and performed the singular value analysis to estimate the largest singular values that quantum mechanics predicts should vanish under ideal conditions. These were found to be on the order of $10^{-1}$ for the real GPT matrix and $10^{-2}$ for the real product GPT matrix, both comparable with the experimental values of $10^{-1.6}$, and $10^{-3.1}$ respectively. These results support our choice of $10^{-1}$ as a well-motivated threshold below which singular values are excluded from the effective rank calculation.\\

\section{Conclusion}
In this experiment, we analyzed data from a two-photon setup within the framework of generalized probability theories to test the principle of tomographic locality. We found no evidence of a violation of tomographic locality at the level of precision of our experiment, 
a conclusion that is consistent with the predictions of quantum theory. Probing further, we purposely restricted the states and effects we targeted in the experiment to a subset that are described by real-amplitude quantum theory, which is theoretically expected to simulate a failure of tomographic locality. This was demonstrated experimentally, where we observed a mismatch between the dimension spanned by full set of such states and effects and the effective dimension spanned by the strictly factorizing ones.

Future work could explore tomographic locality, not at the precision frontier as we have done here, but rather in  experiments that are guided by concrete hypotheses about how the principle might fail in exotic scenarios, which was referred to as the {\em terra nova frontier} in Ref.~\cite{mazurek2021experimentally}.


This work also illustrates how the tensor product structure can be operationally defined through the transformations rather than the preparations or the measurements. Specifically, if, for a bipartite system, all the preparation and measurement procedures are not expected to be described by states and effects that are strictly factorizing, but the transformation procedures {\em are}, then one can impose the constraint of factorization on the representations of the transformations in the fitting procedure and thereby identify a tensor product structure in the real vector space of GPT states. Having a means of reconstructing the tensor product structure from empirical data is critical for many applications. For instance, the entanglement properties of states are only defined relative to a tensor product structure, so that an ability to achieve such a reconstruction is necessary if one wishes to make experimental assessments of these entanglement properties.


\bibliography{localtomoTL}

\section*{Appendix A: Stabilizer States}
The pure stabilizer states of a two-qubit system are defined as the $+1$ eigenstates of maximal commuting sets of tensor products of Pauli operators. The Pauli operators are given by $(I, \pm X, \pm Y, \pm Z)$.

\begin{figure}[!ht]
\includegraphics[width=1\linewidth]{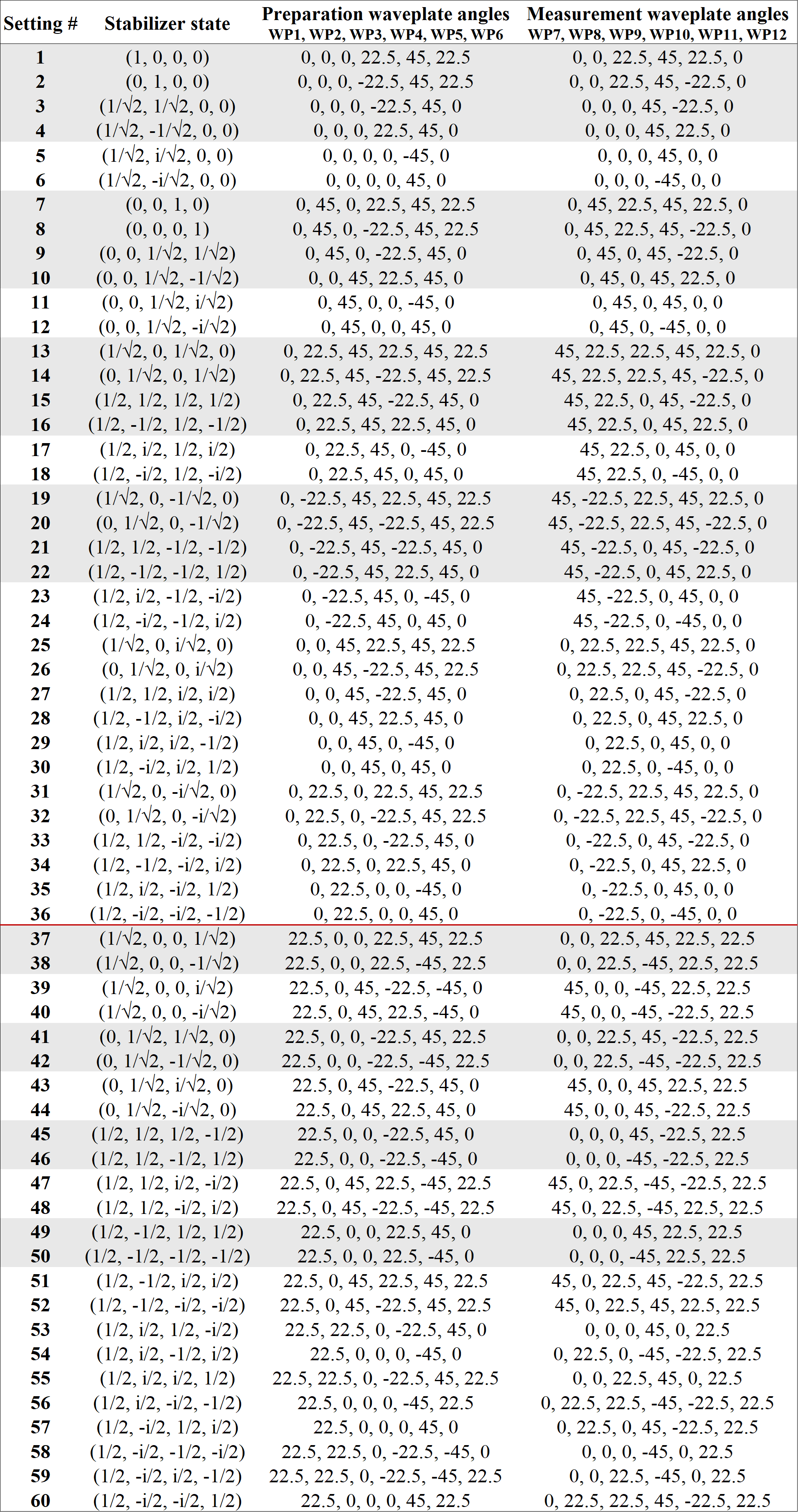}
	\caption{\label{fig:List of measurement}\justifying
	Complete set of 2-qubit pure stabilizer states, 
	 along with the corresponding waveplate angles required for their preparation and measurement on the polarization degrees of freedom of a two-photon system. 
	 States above the red line correspond to product states, while those below it correspond to entangled states. The 2-qubit states that are real-amplitude, hence also 2-rebit states, are highlighted in gray.
	}
\end{figure}

 The product stabilizer states are constructed as $\ket{\psi}_{jk}^{\text{product}} = \ket{\psi}_j \otimes \ket{\psi}_k$, where  $\ket{\psi}_j \in \{\ket{H},\ket{V},\ket{D},\ket{A},\ket{R},\ket{L}\}$. This results in a total of $36$ distinct product states. The entangled stabilizer states are constructed as $\ket{\psi}_{jl}^{\text{entangled}} = \frac{1}{\sqrt{2}}(\ket{H} \otimes \ket{\psi}_j + e^{ i \phi_l}\ket{V}\otimes \ket{\psi}_j^{\perp})$, where $e^{i\phi_l} \in \{+1,-1,+i,-i\}$, and $\ket{\psi}_j^{\perp}$ is the state orthogonal to $\ket{\psi}_j$. This results in a total of $24$ distinct entangled states. A complete list of all 60 stabilizer states (36 product and 24 entangled), expressed in the basis $\{|H\rangle|H\rangle, |H\rangle|V\rangle, |V\rangle|H\rangle, |V\rangle|V\rangle \}$, is provided in the first column of Fig.~\ref{fig:List of measurement}. 
 
For each pure stabilizer state used in the experiment, there is also the corresponding rank-1 stabilizer effect.

\end{document}